\documentclass[%
 preprint,
 superscriptaddress,
nofootinbib,
 amsmath,amssymb,
 aip,
 jap,
 graphicx
]{revtex4-1}

\usepackage{graphicx}
\usepackage{dcolumn}
\usepackage{bm}
\usepackage{hyperref}
\usepackage{tabularx}
\usepackage[version=4]{mhchem}
\usepackage{commath}
\usepackage{color}
\definecolor{lightgray}{gray}{0.75}
\usepackage{booktabs}
\usepackage[capitalise]{cleveref} 
\usepackage[section]{placeins} 
\usepackage{array, makecell} 
\usepackage{siunitx}
\usepackage{pdftexcmds}

\usepackage[myheadings]{fullpage} 
\newcommand\greybox[1]{%
  \vskip\baselineskip%
  \par\noindent\colorbox{lightgray}{%
    \begin{minipage}{\textwidth}#1\end{minipage}%
  }%
  \vskip\baselineskip%
}

\newcommand{\sub}[1]{$_{\text{#1}}$}
\makeatletter
\newcommand\kv[2]{%
  \ifnum\pdf@strcmp{\unexpanded{#1}}{V}=0 %
     \expandafter\@firstoftwo
  \else
    \expandafter\@secondoftwo
  \fi
    {\textit{#1}\!\sub{#2}}
    {#1\sub{#2}}%
}
\makeatother
\makeatletter
\newcommand\kvc[3]{%
  \ifnum\pdf@strcmp{\unexpanded{#1}}{V}=0 %
     \expandafter\@firstoftwo
  \else
    \expandafter\@secondoftwo
  \fi
    {\textit{#1}\!\sub{#2}$^{#3}$}
    {#1\sub{#2}$^{#3}$}
}
\makeatother

\hypersetup{
    colorlinks=true,                
    breaklinks=true,                
    urlcolor= black,                
    linkcolor= blue,                
    bookmarksopen=false,
    filecolor=black,
    citecolor=magenta,
}

\begin{document}

\title{Imperfections are not 0 K: free energy of point defects in crystals}

\author{Irea Mosquera-Lois}
\affiliation{Thomas Young Centre \& Department of Materials, Imperial College London, London SW7 2AZ, UK}

\author{Seán R. Kavanagh}
\affiliation{Thomas Young Centre \& Department of Materials, Imperial College London, London SW7 2AZ, UK}
\affiliation{Thomas Young Centre \& Department of Chemistry, University College London, 20 Gordon Street, London WC1H 0AJ, UK}

\author{Johan Klarbring}
\affiliation{Thomas Young Centre \& Department of Materials, Imperial College London, London SW7 2AZ, UK}
\affiliation{Department of Physics, Chemistry and Biology (IFM), Link\"{o}ping University, SE-581 83, Link\"{o}ping, Sweden}

\author{Kasper Tolborg}
\affiliation{Thomas Young Centre \& Department of Materials, Imperial College London, London SW7 2AZ, UK}
\affiliation{I-X, Imperial College London, London W12 0BZ, UK}

\author{Aron Walsh}
\email{a.walsh@imperial.ac.uk}
\affiliation{Thomas Young Centre \& Department of Materials, Imperial College London, London SW7 2AZ, UK}
\affiliation{Department of Physics, Ewha Womans University, Seoul 03760, Korea}

\date{\today}

\begin{abstract}
Defects determine many important properties and applications of materials, ranging from doping in semiconductors, to conductivity in mixed ionic-electronic conductors used in batteries, to active sites in catalysts. The theoretical description of defect formation in crystals has evolved substantially over the past century. Advances in supercomputing hardware, and the integration of new computational techniques such as machine learning, provide an opportunity to model longer length and time-scales than previously possible. In this Tutorial Review, we cover the description of free energies for defect formation at finite temperatures, including configurational (structural, electronic, spin) and vibrational terms. We discuss challenges in accounting for metastable defect configurations, progress such as machine learning force fields and thermodynamic integration to directly access entropic contributions, and bottlenecks in going beyond the dilute limit of defect formation. Such developments are necessary to support a new era of accurate defect predictions in computational materials chemistry. 
\end{abstract}

\maketitle

\greybox{
{\bf Key learning points}
\begin{itemize}
\item Thermodynamics of point defect formation in crystals
\item Contributions to defect entropies (electronic, spin, vibrational, orientational)
\item Accounting for metastable defect configurations
\item Workflow for calculating defect free energies
\item Outstanding challenges for accurate defect predictions
\end{itemize}
}

\section{Introduction}

The understanding and control of defects in materials are essential for the development of new technologies.
Defects have the power to turn insulating materials conductive, transparent materials coloured, and inert materials reactive. 
In batteries, point defects determine the balance between electronic and ionic transport in solid-state components, and thus charging and degradation rates.\cite{maierThermodynamicsElectrochemicalLithium2013,squiresLowElectronicConductivity2022}
In solar cells, imperfections in the active absorber layer provide recombination pathways for photogenerated electrons and holes that limit efficiency.\cite{shockleyStatisticsRecombinationsHoles1952,kimUpperLimitPhotovoltaic2020}
In photo/electrocatalytic systems, surface layer defects provide active sites that increase reaction rates.\cite{liDefectEngineeringFuel2020,pastorElectronicDefectsMetal2022}
In quantum computers, the spin states of defects can be controlled and measured as a basic unit of quantum information.\cite{weber2010quantum,xiong2023midgap}

The microscopic theory and simulation of point defects in materials have developed over the past century. 
Building on the visionary 1912 work of Born and von Kármán concerning the vibrations of atoms in crystals,\cite{bornUeberSchwingungenRaumgittern1912}  Frenkel considered the thermally activated hopping of an atom from its regular lattice site to an interstitial position in 1926.\cite{Frenkel1926} 
The subsequent `Frenkel pair' formation in a silicon crystal can be described from the defect reaction
\begin{equation}
\textrm{\kv{Si}{Si}} \rightleftharpoons V_{\mathrm{Si}} + \mathrm{Si}_i
\end{equation}
that produces one vacancy \kv{V}{Si} and a corresponding interstitial \kv{Si}{i}\footnote{Defects are often represented as \kvc{M}{s}{c}\!\!~with $M$ being the defect or species occupying the lattice site $s$ and with charge state $c$.}.
The equilibrium fraction of such Frenkel pairs in a sample will depend on the energetic cost of their formation. 
This problem prompted Mott and Littleton to develop a formalism to compute the energies of charged vacancies and interstitials in ionic solids in 1938, which combined an atomistic description of the defect site with a continuum description of the dielectric response of the host crystal.\cite{Mott1938a}
Since then, a variety of defect modelling techniques have been developed including quantum mechanical/molecular mechanical (QM/MM) embedding\cite{grimes1989comparison}, Green's function methods\cite{baraff1984calculation}, and supercell techniques that employ periodic boundary conditions\cite{leslie1985energy}.

In this Tutorial Review, we focus on the free energy of point defect formation at finite temperatures.
The disruption of translational symmetry at a defect site induces changes in the local degrees of freedom, which can be classified into configurational, vibrational, spin and electronic terms. 
We explore these contributions in detail, and describe a modern computational workflow for the systematic calculation of defect free energies.
Finally, we highlight some outstanding computational challenges, including the identification of global minima and the necessary timescales to describe anharmonic potential energy surfaces of defects with accessible metastable configurations. 

\section{Enthalpy-entropy balance}

A crystal in thermal equilibrium at finite temperatures always contains a finite concentration of defects. 
Their formation increases the internal lattice energy of the crystal. Yet this penalty is counterbalanced by an entropy gain, so that the balance between these quantities determines the defect concentration at equilibrium. 
While it has become standard to estimate defect concentrations by calculating their formation energies under constant volume (isochoric) conditions in the absence of temperature, they are actually determined by the Gibbs free energy of the defect reaction, defined under constant pressure (isobaric) conditions at finite temperatures. 
Indeed, the equilibrium {number of defects $n_d$} is determined by minimising the Gibbs free energy of the defective system at constant growth/annealing temperature and pressure:
\begin{equation}\label{eq:1}
    \left( \frac{\partial G_{d}}{\partial n_d} \right)_{P, T, n_{X}} 
    \equiv \frac{\partial G_{d,P}}{\partial n_d} = 0
\end{equation}
{where the total number of atoms $n_X$ of each element $X$ is kept fixed\cite{Morgan_2020,zhang_ab_2022}, and only the pressure constraint is shown for simplicity.} 
$G_{d,P}$ can be separated into two contributions: the free energy of the bulk crystal, $G_{b,P}$, and the change in free energy upon defect formation. The latter is often further decomposed into the configurational entropy $S^{\rm conf}$ and the (non-configurational) free energy $G_{f, P}$ of forming $n_d$ defects at arbitrary lattice sites\cite{Lannoo1981_thermo,Hayes1985,Varotsos1986,Hiroshi2014,Sutton2020,allnatt_lidiard_1993}, such that
\begin{equation}\label{eq:2a}
    \frac{\partial G_{d,P}}{\partial n_d} = 
    \frac{\partial( G_{b, P} + G_{f,P} - T S^\mathrm{conf})}{\partial n_d} =
    \frac{\partial( G_{f,P} - T S^\mathrm{conf})}{\partial n_d}
\end{equation}
In the dilute limit, where there are no defect-defect interactions (i.e. $c=\frac{n_d}{N}<<1\%${, with $N$ being the number of lattice sites where the defect species can form}), \cref{eq:2a} becomes
\begin{equation}\label{eq:2}
    \frac{\partial G_{d,P}}{\partial n_d} = 
    \frac{\partial( n_d g_{f,P} - T S^\mathrm{conf})}{\partial n_d}
\end{equation}
where lowercase letters represent quantities for one defect (e.g. $g_f = \frac{\partial G_f}{\partial n_d}$).

The main driving factor for defect formation is the mixing or configurational entropy $S^\mathrm{conf}$. This arises from the many distinct ways to arrange defects in the solid and can be calculated using
\begin{equation}\label{eq:3}
S^\mathrm{conf} = k_{B} \ln(W)
\end{equation}
with $W$ given by the number of possible arrangements of $n_d$ equivalent defects among the $N$ lattice sites available to that defect species
\begin{equation}\label{eq:4}
W =\ ^{N}\!C_{n_d} = \frac{(N)!}{(N-n_d)!n_d!} \approx \frac{(N)^{n_d}}{n_d!}
\end{equation}
Combining \cref{eq:3,eq:4} and using Stirling's approximation, the configurational entropy is simplified to
\begin{equation}\label{eq:5}
S^\mathrm{conf} = k_B [n_d - n_d \ln(n_d/N)]
\end{equation}
Substituting into \cref{eq:2} and computing the derivative, one obtains
\begin{equation}\label{eq:6}
c_\mathrm{eq} = \frac{n_d}{N} = \exp\left(\frac{- g_{f,P}}{k_B T}\right)
\end{equation}
{where $c_{eq}$ denotes the fraction of available lattice sites $N$ occupied by $n_d$ defects at equilibrium, which can be expressed as a concentration by multiplying by the density of available sites ($[c] = c N/V = n_d / V$, where $V$ is the volume).} In \cref{eq:6}, $g_{f,P}$ is given by
\begin{equation}\label{eq:7}
g_{f,P} = 
h_{f,P} -
T s_{f,P}
\end{equation}
Here, $s_{f,P}$ is the non-configurational entropy contribution per defect, incorporating the changes in all degrees of freedom (spin, orientational, vibrational, electronic, etc.) except for site configurational entropy which was separated from $g_{f,P}$ in \cref{eq:2a}. Combining \cref{eq:6,eq:7} we get
\begin{equation}\label{eq:8}
c_{eq} = 
\exp \left(\frac{s_{f,P}}{k_B}\right) 
\exp \left(\frac{- h_{f,P} }{k_B T}\right)
=
\left( \frac{Z_d}{Z_b}\right) 
\exp \left( \frac{- h_{f,P}}{k_B T} \right) 
\end{equation}
where $\left(\frac{Z_d}{Z_b}\right)$ accounts for the internal degrees of freedom through a ratio of the internal partition function of the defective ($Z_d$) and reference bulk ($Z_b$) crystal\cite{Hayes1985}. This ratio has historically been accounted for with a degeneracy prefactor $g$\cite{Hayes1985,seebauer_charged_2006},
where several approximations can be applied to account for the different internal degrees of freedom, as described in \cref{sec:entropy}. 

\begin{figure}[ht]
    \centering
    \includegraphics[width=0.4\textwidth]{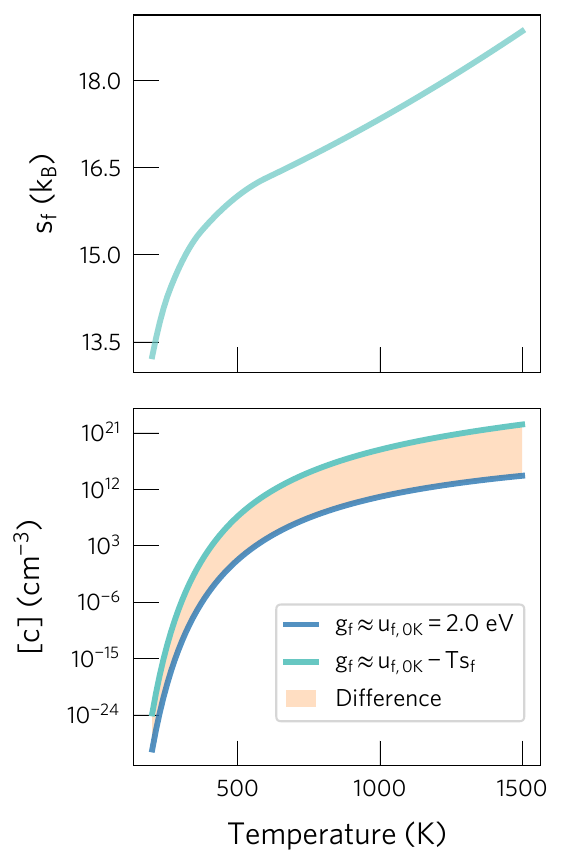}
    \caption{Effect of neglecting entropic contributions when predicting defect concentrations. On the top panel, the defect formation entropy of $\mathrm{O}_\mathrm{i}^{2-}$ in \ce{CeO2} is shown\cite{grieshammer_entropies_2013}. 
    On the bottom one, we show the concentrations predicted when neglecting the formation entropy (dark blue) or including it (light blue).  
    The orange area highlights the error when neglecting entropic effects, which drastically changes predictions at high temperatures. Data adapted from Refs.\citenum{grieshammer_entropies_2013,zacherle_ab_2013}.
    }
    \label{fig:conc}
\end{figure}

If $Z_d$ = $Z_b$ is assumed, one obtains
\begin{equation}\label{eq:9}
c_\mathrm{eq} 
= \exp \left( \frac{- h_{f,P} }{k_B T} \right) 
= \exp \left( \frac{- \left( u_{f,P} + P  v_{f,P} \right)}{k_B T} \right)
\end{equation}
where the elastic term  $Pv_{f,P}$ is negligible at low pressures\cite{Sutton2020,grieshammer_entropies_2013,rauls_entropy_2004,Forslund_2023,lindman_implications_2015} ($Pv_{f,P} \approx 10^{-2}$~meV at an external pressure of \SI{1}{atm} for typical defect formation volumes of $10-20$~\AA$^3$). \cref{eq:9} can then be transformed to the expression widely used in defect studies by applying the approximation $h_{f,P} (T) \approx u_{f,P} (T) \approx u_{f,V} (0~\mathrm{K})$:
\begin{equation}\label{eq:10}
c_\mathrm{eq} = \exp \left(\frac{- u_{f,V}(0~\mathrm{K})}{k_B T} \right) 
\end{equation}
which thus neglects the enthalpic term $Pv_{f,P}$ and finite temperature effects. Yet, the entropic term can be significant at elevated temperatures where many materials are grown or processed and their defects are formed (often 60\%--100\% of the melting point), and thus should be considered for accurate estimations. 
The impact of entropic contributions on predicted defect concentrations is illustrated in \cref{fig:conc}, where neglecting the change in entropy can lead to concentrations underestimated by several orders of magnitude, especially at high temperatures.  
This highlights the importance of a full free energy description when comparing defect concentrations under different growth conditions, since defects with similar formation energies may have different formation entropies, shifting their predicted concentrations at high temperatures\cite{cooper_defect_2018}. 
In the following sections, we describe the different contributions to defect free energies, their relative importance, and how to calculate them.

\section{Defect formation enthalpy}

The defect formation enthalpy can be calculated from the change in internal energy using
\begin{equation}\label{eq:h_f}
    h_{f,P} = u_{f,P} + P v_{f,P}
\end{equation}
where $v_{f,P}$ denotes the change in volume upon defect formation and its enthalpic term is only relevant at high pressures. 
The internal energy change can then be separated into two contributions: a static term and a vibrational one:
\begin{equation}\label{eq:13}
u_{f,P} = u_{f,P}^\mathrm{static} + u_{f,P}^\mathrm{vib}
\end{equation}
The second term is small and often neglected in defect studies, yet can be important for accurate predictions at finite temperatures and will be discussed in \cref{subsec:vib_entropy}. Most defect studies focus on the first term, which is calculated within the supercell framework using\cite{Zhang1991,VandeWalle1993}
\begin{equation}\label{eq:14}
u_{f,P}^\mathrm{static} (T)= U^\mathrm{static}_{d}(V_d) - U^\mathrm{static}_{b}(V_b) - \sum_i{n_i u_i(P, T)} + q (E_F + E_\mathrm{VBM}(V_b)) + E_\mathrm{corr}
\end{equation}
where $U^\mathrm{static}_{b}$ is the potential energy of a supercell of the pristine crystal and $U^\mathrm{static}_d$ of an equivalent supercell containing the defect, and $V_b$ and $V_d$ denote their equilibrium volumes at temperature $T$ and external pressure $P$. The integer $n_i$ indicates the number of atoms of type $i$ that have been added to ($n_i > 0$) or removed from ($n_i < 0$) the supercell to form the defect, and $u_i$ are the corresponding per-atom internal energies of these species (either in their elementary form or competing phases, and at temperature $T$ and pressure $P$). $q$ is the defect charge, $E_F$ the Fermi level or electronic chemical potential {relative to the valence band maximum $E_\mathrm{VBM}$} and $E_\mathrm{corr}$ represents a correction term to account for finite-size effects.\footnote{Note that in metal hosts there are only neutral defects, which simplifies \cref{eq:14} to $u_{f,P}^\mathrm{static} (T)= U^\mathrm{static}_{d}(V_d) - U^\mathrm{static}_{b}(V_b) - \sum_i{n_i u_i(P, T)}$.} As this is the standard approach, with recent reviews focusing on these terms, we refer the reader elsewhere for further information\cite{Freysoldt2014,Kim_2020,Lany_2009b}. We note, however, four important points to consider. 

{Firstly, \cref{eq:13} is often evaluated by calculating the internal energies and the valence band maximum under athermal conditions -- thus neglecting thermal expansion and electron-phonon interactions. Although the effect of temperature on internal energy \emph{differences} is expected to be small, the energies of the band edges have a stronger temperature dependence\cite{Allen_1981,Wickramaratne_2018} -- with band gaps changes on the order of \SI{0.1}{eV} per \SI{100}{K} being typical\cite{monserrat_electronphonon_2018,zhang_temperature-dependent_2020}. This can affect the formation energies of charged defects at growth/annealing temperatures through the Fermi level dependence ($u_{f,P} \propto q(E_F + E_{VBM})$)\cite{qiao_temperature_2022} and hence the predicted defect concentrations. 
While the effect has not been previously explicitly quantified to the best of our knowledge, it can be studied by calculating the band gap at the growth temperature and comparing the predicted concentrations obtained with $E_{VBM}(V_{b,0K}, 0~K)$ and $E_{VBM}(V_{b,T_{growth}}, T_{growth})$.}

Another common approximation in the field involves calculating $u_{f}^\mathrm{static}$ at constant \emph{volume} rather than pressure -- by fixing the volume of the defect supercell to its pristine value -- as discussed in Refs.~\citenum{Catlow1981,Taylor1997a,Varotsos1986}. This is generally a reasonable approximation for $u_{f}^\mathrm{static}$ since the associated error is of the order of \SI{30}{meV}\cite{Freysoldt2014}, and simplifies the application of some of the finite-size corrections\cite{Freysoldt2014} -- which require similar volumes for the pristine and defective cells. However, isobaric conditions are more convenient for a consistent thermodynamic description of the atomic chemical potentials, and thus the constant pressure approach\cite{grabowski_formation_2011} (i.e. volume optimisation of the defect supercell) will be used.\footnote{Finite-size corrections can still be applied by performing the geometry optimisation of the defect supercell in two steps -- first optimising the atomic positions while constraining the shape and volume of the cell, then calculating the finite-size corrections, and finally performing a full optimisation.} 

Secondly, to accurately calculate $U_d^\mathrm{static}$, a stable atomic structure of the defect should be identified. Since structural reconstructions at defects can be significant, a local optimisation of an unperturbed high-symmetry defect configuration often fails to find the ground state, requiring the use of structure searching methods.\cite{Arrigoni2021,Morris2009,Mulroue2011,mosquera-lois_search_2021,Mosquera-Lois2023,shakenbreak2022,kononov_2023,Kavanagh_rapid,wang2023fourelectron, al-mushadani_free-energy_2003} 
A complicating factor is that hybrid non-local exchange-correlational functionals are generally required for the underlying Density Functional Theory (DFT) calculations\cite{du_density_2015,broberg_high-throughput_2023,Deak_2010,Finazzi_2008,Lany_2009,Ganduglia_2009,Clark_2010,Agoston_2009,Janotti_2011,gerosa_accuracy_2017}. 
While local or semi-local exchange-correlation functionals often provide a good approximation for bulk properties, their self-interaction error spuriously disfavours charge localisation, in addition to underestimating the band gap energy. Since the localisation of electrons/holes in both space and energy can result in different defect structures and energies, DFT functionals that accurately 
describe these properties are essential.
{Further, for heavy element systems (period five/six and below), an accurate electronic description \emph{also} involves accounting for relativistic effects like spin-orbit coupling (SOC), which is key to obtaining accurate positions of the band edges and band gap energy -- and thus accurate defect levels and formation energies\cite{du_density_2015,pan_spin-orbit_2018}.\footnote{While the effect of SOC on electronic structure can be significant, its role in geometry relaxation is often small; one pragmatic approach consists of geometry optimisation with scalar-relativistic DFT followed by single-point SOC calculations.} Finally, we note that materials or dopants with highly localised electrons (d/f elements) may require further corrections\cite{chen_nonunique_2022,lee_transition_2022,ivady_role_2013,ivady_first_2018,ivady_theoretical_2014,gerosa_accuracy_2017,walsh_theoretical_2008}.}

Third, while occasionally witnessed in the literature (particularly for semi-local DFT calculations), negative \emph{intrinsic} defect formation energies at the equilibrium Fermi level are typically unphysical, as they indicate that the bulk system would spontaneously decompose through irreversible defect formation. Common causes include an unstable host crystal (e.g. \kv{V}{O} in \ce{KCuO3}\cite{curnan2014effects}) or
a local phase transition triggered by the defect 
(e.g. \kv{V}{O} induced local (tetragonal-like) octahedral rotations in cubic \ce{SrTiO3}\cite{choi2013anti}).
A caveat to this is that a thermodynamically-unstable but kinetically-stable material (e.g. diamond), can exhibit true negative defect formation energies.

Finally, we note that recent advances in finite-size corrections have enabled the calculation of more accurate and reliable defect formation energies\cite{walsh_correcting_2021}. These range from \textit{a posteriori} anisotropic charge corrections\cite{kumagai_electrostatics-based_2014,freysoldt_fully_2009} to self-consistent corrections that directly modify the potential\cite{chagas_da_silva_self-consistent_2021} or charge density\cite{suo_image_2020} in the underlying electronic structure calculation\cite{xiao_realistic_2020}.

\section{Defect formation entropy}\label{sec:entropy}

The defect formation entropy comprises the entropy change upon creating one defect at a specific lattice site. 
As this entropy change is defined for a given site, it does not include the mixing or off-site configurational entropy -- which arises from the multiple ways of placing the defect in \emph{different} lattice sites. It does, however, include the on-site configurational or orientational contribution, which results from inequivalent orientations of the defect at the same site due to a lowering of the local symmetry,
as discussed below. 

Beyond the orientational contribution, spin can also result in an entropy change -- a defect with one unpaired (collinear) electron has two equivalent electronic configurations as the electron can have up or down spin. 
Two additional contributions stem from changes in the vibrational and electronic entropies. The first is mainly determined by changes in the atomic vibrations of the defect environment, while the second stems from changes in the thermal occupation of electronic states. 

\begin{figure}[ht]
    \centering
    \includegraphics[width=1.0\textwidth]{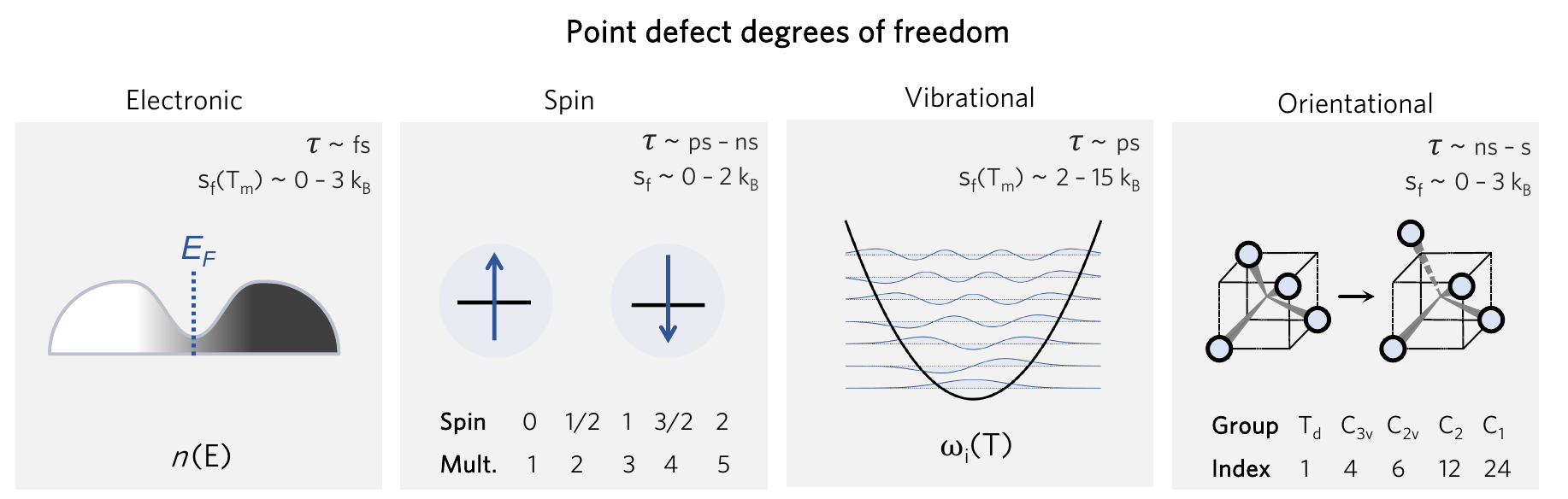}
    \caption{Various degrees of freedom for point defects, with their typical timescales $\boldsymbol{{\tau}}$\cite{ruban_configurational_2008,Yang_2014} and formation entropy ranges $s_f(\textrm{T})$, where $\textrm{T}_{\textrm{m}}$ is the melting temperature.}
    \label{fig:defect_entropies}
\end{figure}

For convenience when calculating defect concentrations, these contributions can be accounted for with the pre-exponential factor $Z_d/Z_b$, as described in \cref{eq:8}. Considering the different timescales of these degrees of freedom (\cref{fig:defect_entropies}), we can often treat them independently and thus express the partition function as a product of the different contributions:
\begin{equation}\label{eq:partitionfunction}
Z = 
Z^\mathrm{electronic} 
Z^\mathrm{spin} 
Z^\mathrm{vibrational} 
Z^\mathrm{orientational} 
\end{equation}
or equivalently,
\begin{equation}
s_f = 
s_f^\mathrm{electronic} +
s_f^\mathrm{spin} + 
s_f^\mathrm{vibrational} +  
s_f^\mathrm{orientational}.
\end{equation}

These entropic terms and the resulting prefactors are illustrated in \cref{fig:defect_entropies} and exemplified for a series of defects and host crystals in \cref{tab:degeneracy_factors}. 
The spin-degree of freedom can be described by $\frac{Z_d^\mathrm{spin}}{Z_b^\mathrm{spin}}=2S+1$, where $S$ is the total spin angular momentum.\footnote{
When SOC is significant, for systems containing heavy elements, the total spin angular momentum should be considered to account for the possible combinations and orientations of the electron spins and their interaction with orbital motion.
}
For example, the unpaired electron ($S=\frac{1}{2}$) for a neutral chlorine vacancy in NaCl results in a prefactor of 2. 
If the orientational degrees of freedom also change, the spin prefactor is multiplied by the orientational one, with the latter determined by the number of symmetry-equivalent orientations of the defect (e.g. 4 for the $C_{3v}$ distorted $V_\mathrm{Cd}^{-1}$, resulting in $\frac{Z_d}{Z_b}=4 (2(\frac{1}{2})+1)=8$). 
Finally, excited states are accounted for in the usual sum over levels $i$ with energies $E$ in $Z = \sum_{i}{e^{-E_i / (k_B T)}}$.
For example, the chlorine divacancy in NaCl shows a sixfold orientational degeneracy, as well as a $S=1$ spin excited state at energy $E$ above the $S=0$ ground state, resulting in the prefactor 
$\frac{Z_d}{Z_b}=6(1+ (2S+1) e^{-E/ (k_BT)})=6(1+ 3 e^{-E/ (k_BT)})$. In the following sections, we describe in detail these degrees of freedom and how to approach the calculation of the associated entropic terms. 

\begin{table}[ht!]
    \centering
    \caption{Defect pre-exponential factors ($Z_d/Z_b$), following Hayes and Stoneham,\cite{Hayes1985} where $Z$ is calculated using $Z = \sum_{i}{e^{-E_i / (k_B T)}}$ and $E_i$ represents the energy of the available states. For simplicity, the vibrational degree of freedom is not included.}
    \label{tab:degeneracy_factors}
    \begin{tabular}{c|c|c|c}
 \hline
 \textbf{Host Crystal} & \textbf{Defect Species} & \textbf{Degrees of Freedom} & \textbf{$\frac{Z_d}{Z_b}$} \\ 
 \hline
 Xe & V$_{\mathrm{Xe}}$--V$_{\mathrm{Xe}}$ & Orientational, $<111>$ & 4 \\
 NaCl & V$_{\mathrm{Cl}}$ & Spin, S=$\frac{1}{2}$ & 2 \\
 NaCl & V$_{\mathrm{Cl}}$--V$_{\mathrm{Cl}}$  & 
 \makecell{(i) Spin, ground state S=0 and excited state \\S=1 at $E$. (ii) Orientational, $<110>$} & $6(1+3e^{-E/ k_BT})$ \\
CdTe & V$_{\mathrm{Cd}}^0$ & Orientational, $C_{2v}$ distortion & 6 \\
CdTe & V$_{\mathrm{Cd}}^{-1}$ &
\makecell{(i) Orientational, $C_{3v}$ distortion. (ii) Spin, S=$\frac{1}{2}$} & 8 \\
CdTe & V$_{\mathrm{Cd}}^{-2}$ & None & 1\\
     \hline
    \end{tabular}
\end{table}

\subsection{Orientational contributions}\label{subsec:orientational_entropy}

\begin{figure}[ht!]
    \centering
    \includegraphics[width=0.8\textwidth]{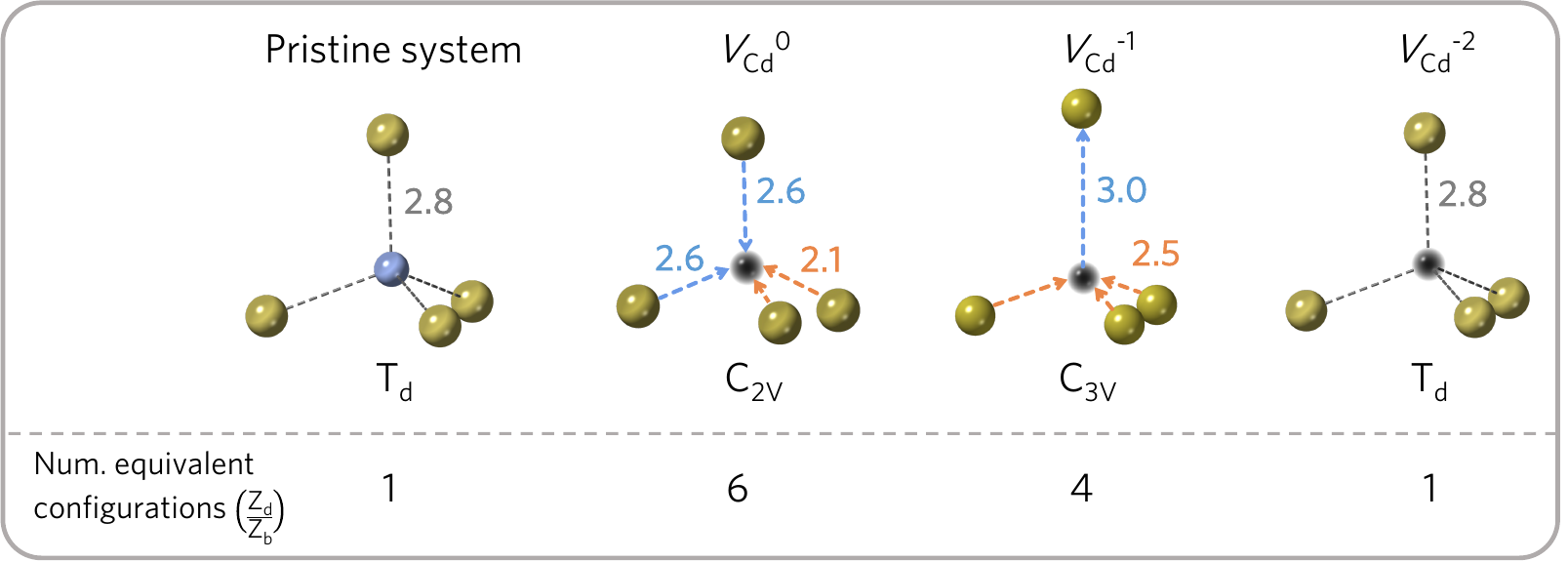}
    \caption{Orientational degrees of freedom for the cadmium vacancy in CdTe. The distortion undergone by the different charge states determines the number of symmetry-equivalent configurations ($Z_d^\mathrm{orient}/Z_b^\mathrm{orient}$). Cd in blue, Te in gold and vacancy in black. For each configuration, different colours are used to group bonds with equal distances (in \AA), illustrating the defect site symmetry. Adapted from Ref.\citenum{Kavanagh_rapid}.}
    \label{fig:orientational}
\end{figure}

The number of equivalent defect orientations can be estimated from the change in point group symmetry at the defect site. 
For example, each Cd site in zincblende-structured CdTe has a $T_d$ point group (with 24 symmetry operations). 
If a \kv{V}{Cd} defect maintains this environment then no additional factor is required. However, a $C_{2v}$ (4 symmetry operations - i.e. order 4) or $C_{3v}$ (order 6) distortion would lead to $\frac{Z_d}{Z_b} = \frac{Z_d}{24} =$ 6 and 4, respectively, as determined by the index of the subgroup\cite{Goede1972,Krasikov2018} and illustrated in \cref{fig:orientational}. 
In practice, this factor can be determined by calculating the ratio of point symmetry operations for the defect site in the pristine and defective supercells using materials analysis codes\cite{Ong2013}. 

The orientational contribution can significantly increase predicted defect concentrations. For instance, a change from $T_d$ to $C_1$ symmetry would result in a 24-fold increase in the predicted concentration. This contribution is also important for calculating the concentration of defect complexes at finite temperatures -- where one has to account for the difference in orientational entropy of the complex and its associated point defects, as well as the loss in off-site configurational entropy, to predict the temperature-dependent binding energy required to overcome entropically-driven dissociation into the constituent point defects\cite{wynn_structures_2016,Krasikov_2017,Millican2022}.

\subsection{Electronic contributions}\label{subsec:electronic_entropy}

The electronic density of states (DOS) can also change upon defect formation. 
As a first approximation for low temperatures, the electronic entropy can be estimated using the Sommerfeld approach\cite{Wolverton1995}. Here, one assumes that the DOS is temperature-independent and varies very slowly for energies near the Fermi level, resulting in
\begin{equation}
    S^\mathrm{elec}(T) = \frac{\pi^3}{3} k_B^2 T D(E_F) 
\end{equation}
where $D(E_F)$ represents the DOS at the Fermi level $E_F$. A more sophisticated treatment involves using the fixed DOS approximation\cite{Eriksson1992}, which only assumes a temperature-independent density,
\begin{equation}\label{eq:elec_entropy}
    S^\mathrm{elec} = 
    - k_B 
    \int_{-\infty}^{\infty} D(E) 
    \left(
        f(E, T) 
        \ln{\left( f(E, T) \right)}
        + 
        \left( 1 - f(E, T) \right) 
        \ln{\left( 1 - f(E, T) \right)}
    \right)
    dE
\end{equation}
where $D(E)$ is the electronic density of states at energy $E$ (calculated at \SI{0}{K}) and $f(E)$ is the occupation of the energy level $E$ given by Fermi-Dirac occupation statistics
\begin{equation}
    f (E, T) = \left( 
    \exp{\left( \frac{E - E_F}{k_B T}\right)} + 1 
    \right) ^{-1}
\end{equation}
Further accuracy requires using self-consistent finite temperature DFT to calculate a temperature-dependent DOS\cite{Mermin_1965}. However, since the effect of temperature on the density of states profile is typically small and generally affects pristine and defective systems in a similar way, the fixed DOS method is often a reasonable approximation and yields electronic entropies in good agreement with more accurate (and computationally expensive) finite temperature DFT approaches\cite{zhang_accurate_2017,satta_vacancy_1998}. 

Generally, changes in the electronic degrees of freedom are only significant for metals or narrow band gap semiconductors at high temperatures\cite{satta_vacancy_1998,youssef_intrinsic_2012}, with absolute values of $s_f^\mathrm{elec}$ ranging from 1 to 3$k_B$. For instance, at the melting point of the corresponding metals, $s_f^\mathrm{elec}$ is $1.7 k_B$ for the tungsten vacancy\cite{satta_vacancy_1998}, $-0.5 k_B$ for the tantalum vacancy\cite{satta_first-principles_1999}, and 
$1.6 k_B$ for the nickel vacancy\cite{metsue_contribution_2014}.

\subsection{Vibrational contributions}\label{subsec:vib_entropy}

\subsubsection{Harmonic treatment}\label{subsec:harm}
Beyond changes in the local atomic arrangement, defects can also modify the vibrations of a crystal. 
A point defect may produce localised vibrations (e.g. 2854--3096 cm$^{-1}$ modes for H in ZnO)\cite{nickel2003hydrogen} and/or perturb the phonon dispersion of the host materials (e.g. a redshift in optical phonon modes by Se$_{\mathrm{S}}$ impurities in ZnS)\cite{dimitrievska2016resonant}.

In early theoretical studies, the vibrational entropy was approximated by \emph{only} considering the change in force constants of the defect's nearest neighbours\cite{Kroger1964,Hiroshi2014}.
With increased computational power, all vibrations can be considered. 
The simplest approach involves applying the harmonic approximation, where the vibrational entropy is calculated by appropriately summing the phonon frequencies $\omega$ over bands $v$ and wavevectors $q$
\begin{equation}
    S^\mathrm{vib} = 
    \frac{1}{2T} 
    \sum_{q, \nu}{
        \hbar \omega_{q, \nu} 
        \coth{\left( 
            \frac{\hbar \omega_{q, \nu}}{2k_B T} 
        \right)}
    }
    - k_B \sum_{q, \nu}{
        \ln  \left ( 
            2 \sinh{\left( \frac{\hbar \omega_{q, \nu}}{2k_B T} \right)}
        \right)
    }
\end{equation}
In practice, this is generally replaced by the vibrational Helmholtz free energy to account for the vibrational internal energy including zero-point motion, giving
\begin{equation}\label{eq:vib_free_energy}
    A^\mathrm{vib} = 
    \frac{1}{2} 
    \sum_{q,\nu}{
        \hbar \omega_{q,\nu}
    }
    +
    k_B T \sum_{q,\nu}{
    \ln \left( 
        1 - \exp{\left(\frac{\hbar \omega_{q,\nu}}{k_B T}\right)}
    \right)}
\end{equation}
The vibrational frequencies are obtained from the interatomic force constant matrix, which can be calculated using either the linear response method\cite{Baroni_2001,Millican2022} or finite displacements\cite{Parlinski1997} with codes such as \texttt{phonopy}\cite{phonopy-phono3py-JPSJ}. 
The harmonic vibrational contribution to defect formation can then be obtained by calculations of the defective and pristine systems using \cref{eq:vib_free_energy} and combining them into
\begin{equation}\label{eq:a_f^vib_vol}
    a_{f,P}^\mathrm{vib}(T) = A_d^\mathrm{vib}(V_{d},T) - A_{b}^\mathrm{vib}(V_{b},T) - \sum_i{n_i a^\mathrm{vib}_i(P,T)}
\end{equation}
where a Legendre transformation can be used to obtain the Gibbs free energy\cite{zhang2018calculating,Varotsos1986},
\begin{equation}\label{eq:g_f^vib_vol}
    g_{f,P}^\mathrm{vib}(T) = A_d^\mathrm{vib}(V_{d},T) - A_{b}^\mathrm{vib}(V_{b},T) 
    + P (V_{d} - V_b)
    - \sum_i{n_i \mu^\mathrm{vib}_i(P,T)} 
\end{equation}
In \cref{eq:a_f^vib_vol,eq:g_f^vib_vol}, $V_b$ and $V_d$ represent the equilibrium volume of the pristine and defective supercells at \SI{0}{K} and pressure $P$. The last term accounts for the per-atom vibrational free energy of the external reservoir that acts as a source or sink for atomic species.
We note that, due to the symmetry lowering induced by the defect formation, evaluating $A_d^\mathrm{ vib}$ is often significantly more computationally demanding than $A_{b}^\mathrm{vib}$. For large supercells, the calculation of $A_d^\mathrm{ vib}$ can be simplified by applying the Combined Dynamic Matrix approximation\cite{shi_ab_2012,zhang_antisite_2023,xiao_anharmonic_2020,wu_unambiguous_2018,shi_comparative_2015,zhang_first-principles_2017,wang_origin_2023,alkauskas_first-principles_2014}, which only calculates the interatomic force constants for the interactions affected by the defect formation. 
In practice, a cut-off radius $R_c \approx 3.2-4$~\AA ~is defined around the defect centre, and the interatomic force constants $F_{i,j}$ are only calculated if at least \emph{one} of the atoms $i$ or $j$ is located within the cut-off distance $R_c$. When both atoms are located outside the cut-off distance, their interatomic force constant is approximated by its value in the pristine supercell, whose evaluation is generally more affordable due to the higher symmetry of the pristine supercell.

The vibrational contribution can significantly affect predicted defect concentrations. 
For example, at the melting point, $s_f^\mathrm{vib}$  for the vacancy in elemental Cu, C, and Si account for $4.4~k_B$\cite{glensk_breakdown_2014}, $13.1~k_B$\cite{Fatomi2022} and $9.1-11.6~k_B$\cite{Sholihun2015,al-mushadani_free-energy_2003} ($T_\mathrm{m}=$~\SI{1360}{K}, \SI{4100}{K} and \SI{1685}{K}), respectively, thus increasing the predicted concentration by factors of $10^2$, $10^{6}$ and $10^{4}-10^{5}$. 
Similarly, for \kv{V}{In} in \ce{In2O3}, the Frenkel defects in \ce{ThO2} and \ce{CeO2}, and \kv{V}{Ga} in \ce{Ga2O3}, accounting for vibrations results in a $10^{2}$, $10^{2}$, $10^{5}$ and $10^{6}$ increase in the calculated concentrations at \SI{1000}{K}, respectively (with $ s_{f}^\mathrm{vib} = 5.3~k_B$\cite{agoston_formation_2009}, $4.5~k_B$\cite{Moxon_2022}, $11.7~k_B$\cite{grieshammer_entropies_2013} and $14~k_B$\cite{zacherle_ab_2013}). 

\subsubsection{Quasiharmonic treatment}\label{subsec:quasi_harm}
The thermal expansion of a crystal is an anharmonic effect that can be accounted for in the quasiharmonic approximation. While still assuming non-interacting phonons, this formalism accounts for the volume dependence of the phonon frequencies - and thus, indirectly, for their temperature dependence. In practice, this involves repeating the harmonic force constant calculation for a range of slightly expanded and contracted lattice constants, so that the \emph{total} free energy (i.e. $U^\mathrm{static} + A^\mathrm{vib}$) can be minimised at different temperatures, as done for instance in Ref.~\citenum{Taylor1997a}. 
{We can then evaluate $u^{static}_{f,P}$ and $g_{f,P}^\mathrm{vib}$ (\cref{eq:g_f^vib_vol,eq:14}) using the equilibrium volumes of the pristine and defective supercells at the target temperature $T$ and pressure $P$ rather than using the athermal volumes.}

Generally, the (quasi)harmonic approximation works well at low to moderately high temperatures. 
However, at high temperatures where the atomic vibrations significantly deviate from the perfect lattice positions, approaching a phase transition or the melting point, higher-order anharmonic effects should be included and will be described in \cref{subsubsec:anharmonic}.

\subsubsection{Anharmonic treatments}\label{subsubsec:anharmonic}
The (quasi)harmonic approximation works well for modelling the thermal physics of many materials but fails when the potential energy surface becomes too complex to describe with quadratic fits. This is the case for systems that exhibit second-order phase transitions involving soft phonon modes or order-disorder phase transitions involving a multi-valley potential energy surface. In general, at elevated temperatures, higher-order anharmonic terms are necessary for accurate free energy predictions\cite{tolborg2022free,zhang2018calculating} and can be included with two approaches: thermodynamic integration or anharmonic phonon theory.

\textit{Thermodynamic integration (TI).}
Molecular dynamics (MD) simulations offer an alternative to a lattice dynamics expansion of interatomic force constants. In principle all orders of anharmonicity can be included, and TI provides a straightforward way of computing the anharmonic free energy.
The most common type of TI is where integration is performed along a coupling parameter, $\lambda$, from a reference system for which the free energy can be calculated analytically -- often the (quasi) harmonic reference -- to the fully anharmonic system. Thus, a set of MD simulations are performed with Hamiltonians $H(\lambda) = (1-\lambda) H^{\mathrm{qh}} + \lambda H^{\mathrm{DFT}}$, where $H^{\mathrm{qh}}$ and $H^{\mathrm{DFT}}$ are the Hamiltonians of the quasiharmonic reference and the full DFT system, respectively.  The anharmonic correction to the (quasi) harmonic free energy then becomes
\begin{equation}
\label{eq:TI}
    \Delta A = \int_0^1 \frac{\partial A}{\partial \lambda} d\lambda = \\
    \int_0^1 \left< \frac{\partial U_\lambda}{\partial \lambda} \right>_\lambda d\lambda = \\
    \int_0^1 \left< U^{\mathrm{DFT}} - U^{\mathrm{qh}} \right>_\lambda d\lambda ,
\end{equation}
where $\left< \cdots \right>_\lambda$ indicates an average over an MD simulation with Hamiltonian $H(\lambda)$. To compute the anharmonic correction to the free energy of defect formation, (quasi)harmonic phonon calculations should thus be performed for both the bulk and defective systems, and separate TIs (\cref{eq:TI}) then need to be evaluated on top of the harmonic Hamiltonians.   

The TI is most conveniently performed in the NVT ensemble, thus yielding the correction to the anharmonic free energy at constant volume, but constant pressure quantities can be obtained if the simulation cell volume is first determined at the relevant temperature and pressure, using e.g. NPT dynamics. Alternatively, a mapping from the Helmholtz free energy to the Gibbs free energy can be performed following the method of Cheng \& Ceriotti.\cite{cheng_computing_2018}

TI can be performed from a quasiharmonic reference at \emph{each} temperature of interest. 
However, beyond being computationally demanding, performing TI at high temperatures can be challenging due to diffusion or dynamic disorder.
These issues can be avoided by carrying out the coupling constant TI only at a sufficiently low temperature, and then performing TI using temperature as the external coupling parameter to obtain the free energy as a function of temperature, i.e.
\begin{equation}
    \frac{A(V,T_1)}{k_B T_1} - \frac{A(V,T_0)}{k_B T_0} = \\
    - \int_{T_0}^{T_1} \frac{\left< U\right >_T}{k_B T^2} dT ,
\end{equation}
for the Helmholtz free energy, and similarly for the Gibbs free energy at constant pressure by replacing the internal energy, $U$, with the enthalpy, $H$.\cite{cheng_computing_2018}

Since TI using ab initio MD (AIMD) can be computationally expensive due to the extensive sampling required, several methods have been devised for improving convergence, such as the UP-TILD\cite{Grabowski2009} (TU-TILD\cite{duff2015improved}) method; (two-stage) upsampled TI using Langevin dynamics. In the former, TI is performed between the quasiharmonic reference and a poorly converged, but faster, DFT method followed by thermodynamic perturbation from the low-quality DFT to high-quality DFT, while in the latter, an empirical potential is introduced as another stage in the TI to speed up the simulations even further. In both cases, one takes advantage of the fact that low-quality DFT and even empirical potential MD trajectories tend to sample the relevant phase space very well. While this has been shown to be efficient for metals and refractory materials, it is unclear if sampling with low-quality DFT or an empirical potential will be sufficient for defects with more complex electronic structure, such as competing localised and delocalised states.\cite{Kavanagh_rapid,kononov_2023}

The importance of anharmonic contributions to the free energy of defect formation has been shown for vacancies in several metals including Cu, Al and Fe.\cite{glensk_breakdown_2014,cheng_computing_2018}
Despite progress in improved sampling methods, AIMD may still be prohibitively expensive for more complex systems than simple metals -- especially when considering the high accuracy required in the target formation free energy ($\approx$\SI{1}{meV/atom}). Thus, there is promise in using machine-learned force fields (MLFF) (also called interatomic potentials) to reduce the computational cost for accurate thermodynamic calculations. MLFFs have been trained and applied for vacancies in Al, Fe and W, showing good agreement with DFT calculations and experiments.\cite{bochkarev_anharmonic_2019,Goryaeva_2021}
For Al, important differences between MLFFs and DFT were observed in the entropy as a function of temperature, suggesting that the DFT results may be undersampled.\cite{bochkarev_anharmonic_2019} To improve accuracy, one can perform thermodynamic perturbation from the MLFF reference to the full DFT results.\cite{cheng2019ab} 

MLFFs further hold the promise of studying the thermodynamics of defects in more complex systems beyond simple metals, where hybrid DFT or more accurate electronic structure methods are necessary even for a qualitative understanding of the defect formation. While AIMD with hybrid functionals is prohibitively expensive, MLFFs trained to hybrid DFT accuracy are within reach.

\textit{Anharmonic phonon theory.}
Anharmonic lattice dynamics offers an alternative framework, that may overcome potential TI issues including (under-)sampling, finite-size effects, and assumptions of classical dynamics. 
The last decade has shown tremendous progress with methods such as temperature-dependent effective potential (TDEP)\cite{hellman2011lattice}, self-consistent phonon theory (SPCH)\cite{tadano2018first} and stochastic self-consistent harmonic approximation (SSCHA)\cite{monacelli2021stochastic} now being mature techniques with efficient, open source implementations.
These methods have shown great results for temperature-dependent phonon spectra, phase transition temperatures in soft-mode driven transitions, and thermal conductivities, but they also allow for the calculation of anharmonic free energies.\cite{hellman2013temperature,oba2019first}
To our knowledge, these methods have not been applied to study vibrational contributions to defect formation, but it should be straightforward, although at an increased computational cost compared to using the (quasi)harmonic approximation.

\section{Metastable configurations}\label{sec:metastable}

So far we have focused on calculating free energies for the ground state configuration of a defect species. 
However, defects often feature metastable states or local minima in their potential energy surface\cite{Arrigoni2021,Mosquera-Lois2023,Kavanagh_metastable,Kavanagh_rapid,kononov_2023,Cen_2023}, and these will affect the associated free energy of formation.   
At finite temperatures under equilibrium, each configuration will have a population determined by its relative free energy, with transition rates between defect states determined by the corresponding free energy barriers. 
With this perspective, one may calculate the formation free energy of each configuration $i$ and combine them to get the total defect population\footnote{
Here we assume that the different configurations of a defect have the same contribution to the configurational entropy (i.e. different configurations involve the \emph{same type} of lattice site). To consider defects that involve \emph{different} lattice sites (e.g. $T_d$ and $O_h$ for interstitials in cubic crystals), their respective concentrations should be calculated independently using \cref{eq:6}.
}
\begin{align}\label{eq:c_metastable_1}
    c = \sum_{i=0}^{n} c_i 
    = \sum_{i=0}^{n} \exp\left(\frac{-g_{f,P,i}}{k_B T}\right)
    &= \sum_{i=0}^{n} \exp\left(\frac{s_{f,P,i}}{k_B}\right) \exp\left(\frac{-h_{f,P,i}}{k_B T}\right) \nonumber \\
    &= \sum_{i=0}^{n} \frac{Z_i}{Z_b} \exp\left(\frac{-h_{f,P,i}}{k_B T}\right)
\end{align}
where $Z_i$, $s_{f,P,i}$ and $h_{f,P,i}$ denote the partition function, entropy and enthalpy of formation for configuration $i$, and $n$ the number of \emph{accessible} configurations.  
\cref{eq:c_metastable_1} can be refactored to an expression similar to that used in \cref{tab:degeneracy_factors} for excited spin states,
\begin{equation}\label{eq:c_metastable_2}
    c = \exp\left(\frac{-h_{f,P,0}}{k_B T}\right) \sum_{i=0}^{n} \frac{Z_i}{Z_b} \exp\left(\frac{-\Delta h_{f,P,i}}{k_B T}\right) 
\end{equation}
where $h_{f,P,0}$ is the enthalpy of formation of the defect ground state and $\Delta h_{f,P,i}$ are the relative enthalpies of the metastable configurations ($\Delta h_{f,P,i} = h_{f,P,i} - h_{f,P,0}$). 
With \cref{eq:c_metastable_2} we can evaluate the effect of metastable states on defect concentrations. 
For medium to high relative energies ($\Delta h_{f,P} > 0.4$~eV), the effect is negligible -- e.g. for $\Delta h_{f,P} = 0.4$~eV and $T=$~\SI{1000}{K} we get a prefactor $\frac{Z_{0}}{Z_b} + 0.01 \frac{Z_1}{Z_b}$. 
However, this contribution will be significant for defects with (many) low-energy metastable configurations ($\Delta h_{f,P,i} < 0.1$~eV), especially if coupled with large formation entropies ($Z_i >> Z_b$). For instance, a metastable state with $\Delta h_{f,P} = 50$~meV, $s_{f,P}^\mathrm{vib} = 5~k_B$, and $S=1$ at \SI{1000}{K}, we get a prefactor $\frac{Z_{0}}{Z_b} + 0.5 \frac{Z_1}{Z_b} = \frac{Z_{0}}{Z_b} + 250$, so that the predicted concentration may be increased by up to a factor of 250.
We note here that there are several \emph{non-equilibrium} cases where metastable states are also important, such as in solar cells under illumination\cite{Kavanagh_rapid,Kavanagh_metastable,Zhang_2023,Krasikov2018} or materials under electric fields/irradiation\cite{Ewels_2003}, where metastable defect populations can be greatly increased due to kinetic effects. Moreover, metastable states make up the intermediate configurations in defect migration trajectories, so accurately modelling their associated energy surfaces is of key importance for predicting ionic conductivities and kinetic decomposition.

\begin{figure}[ht]
    \centering
    \includegraphics[width=0.9\textwidth]{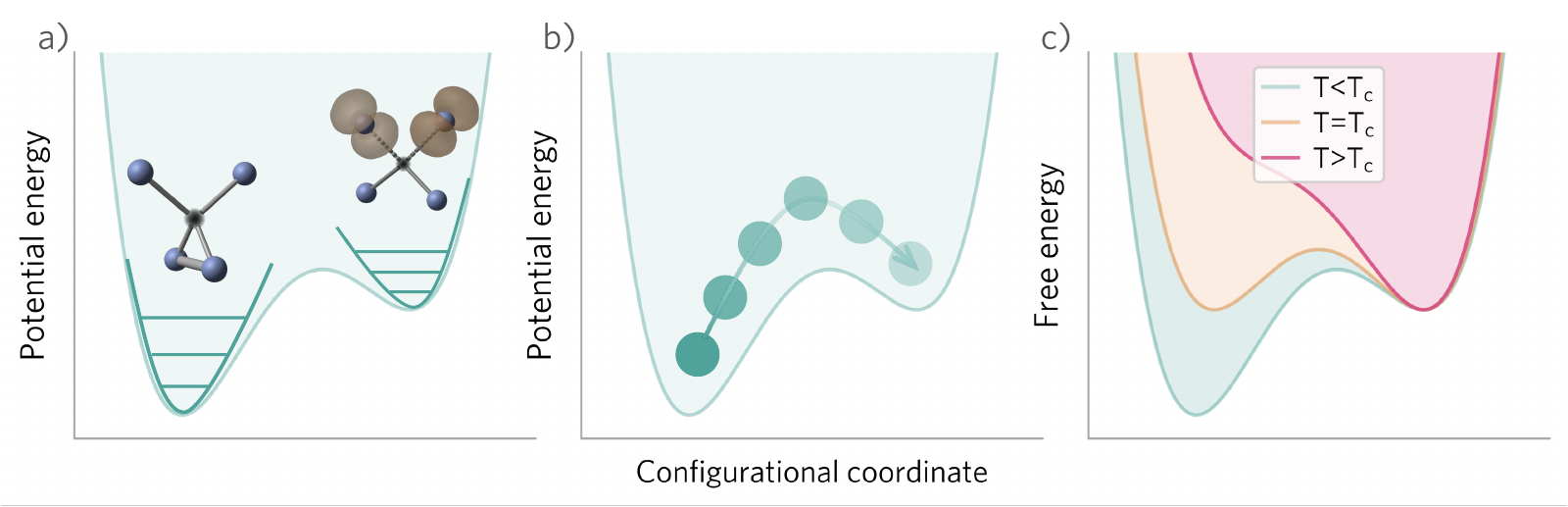}
    \caption{Potential and free energy landscape for a defect with a low-energy metastable configuration. 
    a)~Separation of configurational and vibrational degrees of freedom, so that the free energy may be calculated independently for each configuration (e.g. using harmonic or anharmonic phonon methods for the vibrational free energy). 
    b)~Transition between defect configurations, exemplifying how thermodynamic integration could be used to sample over the thermally accessible structures, thereby calculating the total free energy of the defect.   
    c)~Example of a defect whose ground state structure changes with temperature. $T_c$ denotes the critical temperature where the free energies of both structures are equal.}
    \label{fig:metastable}
\end{figure}

In these cases, defects will exhibit strong vibrational anharmonicity due to the presence of thermally accessible metastable configurations on their energy landscapes, likely requiring anharmonic treatments for accurate predictions. 
A TI approach may be especially applicable, as it could allow one to sample over all possible configurations of a particular type of defect, thus obtaining the total defect free energy without needing the abstraction into vibrational and orientational degrees of freedom, as illustrated in \cref{fig:metastable} (a, b).
This will be important when the time-scale of vibrations within each local minimum and jumps between them becomes similar. Furthermore, if the defect structure changes with temperature, as exemplified in \cref{fig:metastable} (c), the total free energy will naturally emerge from this method without having to consider a possible change of the orientational entropy with temperature. Similarly, defect migration would be handled naturally by this method.

In contrast, if the \emph{barriers} between configurations are large, giving rise to slow and rare transitions that would not be adequately sampled in standard MD ensembles, a TI approach would be less applicable. 
In these cases, one could use harmonic (or anharmonic) phonon-based methods to calculate the (an)harmonic vibrational free energy for the relevant defect configurations, as illustrated in \cref{fig:metastable} (a). By yielding the temperature-dependent \emph{free} energy surfaces of defective systems, these approaches allow for an understanding of defect structure as a function of temperature (\cref{fig:metastable} (c)), where one may envision that thermal effects -- or nuclear quantum effects -- could increase the symmetry of symmetry-broken defects, as occurs for symmetry-broken bulk materials.\cite{klarbring2020anharmonicity,errea2020quantum}

\section{Defect free energy workflow}\label{sec:free_energy}

Drawing together the concepts discussed above, we describe how to combine these contributions for the systematic calculation of defect free energies.

\subsection{Enthalpy change}

The enthalpy change upon defect formation is usually the dominant contribution and an important starting point. The first step is to construct a pristine supercell, whose size should be large enough to minimize defect-defect short-range interactions (i.e. lattice parameters $>10$~\AA\cite{Freysoldt2014,Lany_2008}). The defect structure can then be generated by adding or removing the corresponding atom(s) from the pristine supercell -- which can be automated 
using packages such as \texttt{PyCDT}\cite{broberg_pycdt_2018}, \texttt{PyLada}\cite{goyal_computational_2017}, \texttt{DASP}\cite{huang_dasp_2022}, \texttt{Spinney}\cite{arrigoni_spinney_2021}, \texttt{pymatgen-analysis-defects}\cite{jimmy_shen_2023_7803614} or \texttt{doped}\cite{doped}.
Once the defect supercell has been created, a structure searching method\cite{Mosquera-Lois2023,Arrigoni2021} should be employed to identify stable structures for each of the relevant charge states -- as implemented in \texttt{ShakeNBreak}\cite{shakenbreak2022} for example. 

From the local minimum configurations identified, one should decide whether their effect on the defect thermodynamics is likely to be relevant (depending on their relative energy and entropy). Different criteria will apply in \emph{non-equilibrium} cases.\cite{Kavanagh_metastable,Krasikov2018} For simplicity, here we assume that only the ground state structure is important, yet the procedure can be easily adapted to include metastability (see \cref{sec:metastable}). 
Next, the defect enthalpy of formation can be calculated for the ground state structure using \cref{eq:14,eq:h_f}. 
This requires calculating the finite-size corrections using one of the available methods\cite{freysoldt_fully_2009,kumagai_electrostatics-based_2014,xiao_realistic_2020,suo_image_2020}, as well as the chemical potentials of the relevant competing phases -- steps that can be performed with the mentioned packages. 

\subsection{Entropy change}

We next consider entropic contributions. 
Starting with the temperature-independent terms, the spin degeneracy is determined by the number of unpaired electrons (e.g. from $\frac{Z_d^\mathrm{spin}}{Z_b^\mathrm{spin}}=2S+1$), while the orientational degeneracy can be estimated from the change in symmetry of the defect site using codes such as \texttt{pymatgen}\cite{Ong2013} (\cref{subsec:orientational_entropy}). 
Within host materials that exhibit various types of disorder, such as site\cite{Huang2022}, spin\cite{Oleksandr_2023}, rotational, polar\cite{Choi_2013,Zhang_2006} or elastic/structural\cite{Oleksandr_2023}, the introduction of defect species can lead to a reduction in configurational entropy due to short-range ordering induced by the defect, which should be accounted for if relevant.
If modelling a metal or a narrow band gap semiconductor, then the (temperature-dependent) electronic contribution should be included, which can be estimated using \cref{eq:elec_entropy} with the calculated density of states for the pristine and defective supercells (\cref{subsec:electronic_entropy}).

The vibrational contribution can be calculated using an appropriate approximation (\cref{subsec:vib_entropy}), depending on the target system and temperature, desired accuracy and available resources. 
At low temperatures, vibrational contributions may be negligible, but at high operating or annealing temperatures, their effect on free energies will be significant (\cref{subsec:vib_entropy}).
One should consider whether the harmonic approximation reasonably describes the target system at the conditions of interest. This involves assessing the importance of thermal expansion {and anharmonic effects} for the target system and temperature (e.g. by considering the magnitude of the volumetric thermal expansion coefficient and the existence of low-lying metastable states{, respectively}). 

Based on these considerations, we highlight three possible approaches.   
The simplest is to employ the harmonic approximation to calculate the (temperature-dependent) vibrational entropy for the defective and pristine supercells at the \SI{0}{K} volume, using \texttt{phonopy}\cite{phonopy,phonopy-phono3py-JPSJ} (\cref{subsec:harm}). 
Alternatively, thermal expansion can be accounted for using the quasiharmonic approach to calculate the vibrational entropy for a range of cell volumes for the defective and pristine supercells (\cref{subsec:quasi_harm}). 
Finally, if higher-order anharmonic effects are expected to be important, one can use thermodynamic integration or anharmonic phonon-based methods (\cref{subsubsec:anharmonic}). However, their associated computational cost may prevent their application in certain cases. Here, reverting to the (quasi)harmonic approximation to include vibrational contributions will already be a significant improvement upon static methods that completely neglect finite temperature effects.

\subsection{Chemical potentials}
When defect formation involves the exchange of atoms with an external reservoir, it is essential that consistent thermodynamic potentials are employed throughout. For the special case of stoichiometric defect formation, including Frenkel and Schottky pairs, the chemical potential terms cancel out and these complications can be avoided. 

The atomic chemical potentials in \cref{eq:14} must be modified to contain the same entropic contributions as considered for the defective and pristine systems, e.g.
\begin{equation}
\mu (P,T) = u(P, T) + Pv - T(s^\mathrm{spin} +  s^\mathrm{vib} + s^\mathrm{elec} + s^\mathrm{rot} + s^\mathrm{gas/liquid})
\end{equation}
where lowercase letters are used since these are per-atom quantities and the internal energy $u(P,T)$ includes vibrational contributions (zero-point motion and heat capacity terms). The dependence of the entropy terms is not shown for simplicity. 

For gaseous/liquid reference phases,  rotational ($s^\mathrm{rot}$) and translational/configurational ($s^\mathrm{gas/liquid}$) entropy contributions can become significant. For gases, these quantities may be calculated analytically for arbitrary partial pressures via the ideal gas and rigid rotor approximations\cite{Jackson2016,ase-paper,vaspkit}. 
For liquids or non-ideal gases on the other hand, the thermodynamic potentials can be obtained from molecular dynamics trajectories\cite{Zhang2011} or taken from standardised tables of experimental data\cite{nist}. 
These adjustments are necessary to ensure a consistent thermodynamic description when species are added or removed from the system. 
Chemical potential terms allow specific growth conditions to  be considered, through the choice of temperature/pressure or a tabulated reference potential. 

Determining the chemical potential limits by considering the free energy of all possible secondary phases would involve a significant increase in computational cost.\cite{cplap} A reasonable approximation here is to query a materials database to search for the phases which border the host material on the phase diagram, or are within a given energy error threshold of bordering the host, calculate their internal energy with the appropriate computational setup and then determine the relevant (nearly-)bordering competing phases based on these energies, keeping in mind that entropy contributions will be larger for gases/liquids.\cite{doped} The entropic terms can then be calculated for only these competing phases that directly limit the stability region of the host.

\subsection{Fermi level}
After combining all terms into the defect formation free energy, the remaining variable is the Fermi level, $E_F$.
The position of $E_F$ is dictated by the condition of net charge neutrality, such that the sum of all excess positive charge in the material (from donor-type defects and holes) equals that of all excess negative charge (from acceptor defects and electrons).
As the energies and thus concentrations of charged defects themselves depend on the Fermi level, with $g_{f,P} (T) = (...) + q (E_{VBM} + E_F(T)) - T s^\mathrm{elec}(E_F)$ where $q$ is the defect charge, both $E_F(T)$ and the defect free energies and concentrations must be solved self-consistently under the net neutrality constraint\cite{scfermi,Squires2023}.

{Typically materials are grown or processed/annealed under elevated temperatures where defects form with concentrations given by \cref{eq:6}, before the material is cooled (`quenched') to the operating temperature. While the \emph{total} concentration of each defect is kept fixed due to kinetic trapping, the Fermi level and thus relative populations of different charge states and electron/hole carrier concentrations re-equilibrate upon cooling.
This is modelled by calculating the formation free energies at the growth/annealing temperature, which gives the \emph{total} concentration of each defect that forms during synthesis (using \cref{eq:6}). To then calculate the Fermi level, free carrier concentrations and relative concentrations of the different charged defects at the \emph{operating} temperature, these are solved self-consistently while fixing the total concentration of each defect\cite{scfermi,Squires2023,huang_dasp_2022,Nicolson_2023,Yang_2014,villa_role_2022,shousha_tuning_2020}}
\footnote{On the other hand, if the material is cooled sufficiently slowly from the growth to the operating temperature so that thermodynamic equilibrium can be assumed, then the defect concentrations should be calculated at the operating temperature without fixing the total defect populations.\cite{scfermi,Yang_2014}}. The population of each defect with charge $q$ at the operating temperature $T$ is then given by
\begin{equation}
    c_{d^q} = c_d \frac{\exp \left(\frac{- g^q_{f,P}(E_F; T)}{k_B T} \right)}
    {\sum \limits_{q} \exp \left(\frac{- g^q_{f,P}(E_F; T)}{k_B T} \right)}
\end{equation}
where $c_d$ is the total concentration of defect $d$ (calculated at the growth/annealing temperature) and $g^q_{f,P}$ is the formation free energy in the $q$ charge state.
As mentioned, in solving for the (self-consistent) Fermi level, the electron and hole carrier concentrations are also computed, giving the predicted doping behaviour.
The calculated defect concentrations can then be used in the prediction of a range of defect-related material properties, including carrier recombination rates,\cite{kimUpperLimitPhotovoltaic2020,Kavanagh_rapid,Kavanagh_metastable} ionic/electronic conductivity balance,\cite{maierThermodynamicsElectrochemicalLithium2013,squiresLowElectronicConductivity2022} catalytic activity\cite{liDefectEngineeringFuel2020,pastorElectronicDefectsMetal2022} and or any other extensive defect properties.

\section{Challenges and Outlooks}

The calculation of defect free energies poses a significant computational challenge, especially when considering all intrinsic defects in structurally or chemically complex systems. This complexity can lead to many inequivalent defect species that need to be considered\cite{Zhang_2023}. Depending on the application, one can envision different strategies to make the problem more tractable. 

One approach would be to filter the configurational space and only calculate the entropic terms for those defects with the lowest formation energies and thus highest concentrations.
Alternatively, one could employ reasonable approximations to make the calculation of entropic terms more efficient. For instance, this could involve employing (semi-)local DFT exchange-correlation functionals for calculating vibrational entropies, instead of their hybrid counterparts often used for the energetic terms. Considering that (semi-)local functionals generally describe pristine structures and force constants accurately, this may be a reasonable approximation \emph{if} the defect structure is stable at that level of theory.

Alternatively, a surrogate model could be used instead of first-principles methods.
Indeed, much of the existing defect literature has been built on the development and application of classical force fields.
Given the remarkable progress achieved in machine learning force fields (MLFFs), which can provide a more flexible and accurate description of the potential energy surface,\cite{george2020combining,deringer2021gaussian} they may be an optimal solution in certain cases. 
For instance, MLFFs may be appropriate when targeting in-depth studies of specific defects (e.g. metastable states or migration paths) or requiring high accuracies (e.g. including anharmonic interactions\cite{Goryaeva_2021,bochkarev_anharmonic_2019,Forslund_2023}). 
The significant cost of these calculations would justify training a model for the defective systems, which can be achieved via fine-tuning (e.g. training a model for the bulk and then re-training it for relevant configurations of the defective systems\cite{pols_how_2023}). 
We highlight that further work is still required to determine the optimal approach for defect MLFF training, particularly regarding the number and diversity of defect configurations required to achieve sufficient accuracies. 
Additional research is also needed to investigate whether current MLFFs can describe defects with complex electronic structures and configurational landscapes, a challenge that will likely benefit from progress in fourth-generation MLFFs that include local charges and non-local effects\cite{ko_general-purpose_2021}.  

Beyond using surrogate models to reduce computational cost, other challenges include going beyond the dilute non-interacting limit that is often assumed and thus has been our focus ($c << 1 \%$).
At higher concentrations, interactions between defects must be considered. These can modify both the internal energy and the accessible degrees of freedom. Inspired by electrolyte models, Debye-Hückel theory has been used to account for long-range Coulombic interactions in certain highly defective ionic systems.\cite{fong1969maxwell} Here the Coulomb interaction is modified by a screening term that depends on the concentration of charged defects, which can be formulated in terms of an activity coefficient. 
Future models could extend to account for the additional effects on enthalpies (e.g. elastic and bonding) and entropies arising from finite defect separations.

At shorter length scales, defect complexation can also occur, such as the combination of a vacancy and interstitial to form a bound Frenkel pair. Defect aggregation will alter the configurational landscape, in addition to vibrational, electronic, and spin terms.\cite{kroger1977defect}  Defect complexes have been characterised in many host compounds including the NV centre in diamond.\cite{lenef1996electronic} However, a complete description of the full ensemble of configurations that can be formed is challenging beyond certain high symmetry cases. 
One solution is a combinatorial evaluation of defect configurations in a radial cluster expansion, as described by Allnat and Loftus.\cite{allnatt1973physical}. At higher concentrations, the description of long-range ordering may be required with the emergence of new non-stoichiometric phases.\cite{tomlinson1990computer,Sopiha_2022} 

These cases illustrate where further developments are required for accurate defect predictions. 
Surrogate models will enable accessing longer time and length-scales, required to describe anharmonic effects and go beyond the dilute non-interacting limit.
In this Tutorial Review, we have described the different terms that contribute to defect free energies and how to calculate them. We have highlighted the importance of including entropic effects for accurate defect concentrations, and described reasonable approximations to reduce the associated computational costs. Lastly, we have discussed the remaining challenges and potential solutions for comprehensive modelling of defect free energies.

\acknowledgments
I.M.L. acknowledges Imperial College London for funding a President’s
PhD scholarship. 
S.R.K. acknowledges the EPSRC Centre for Doctoral Training in the Advanced Characterisation of Materials (CDT-ACM)(EP/S023259/1) for funding a PhD studentship. 
K. T. acknowledges support from the Eric and Wendy Schmidt AI in Science Postdoctoral Fellowship, a Schmidt Futures program.
J. K. acknowledges support from the Swedish Research Council (VR) program 2021-00486.
A. W. acknowledges support from the  Leverhulme Trust.

\bibliographystyle{rsc}
\bibliography{defects,extra}

\end{document}